\documentclass[aps,prb,twocolumn,floats]{revtex4}

\usepackage{amssymb}
\usepackage{amsmath}
\usepackage{graphicx}
\usepackage{bm}
\usepackage{mdframed}

\newcommand{\ud}{\,\mathrm{d}}

\begin{document}

\bibliographystyle{prsty}
\author{H\'{e}ctor Ochoa, Ricardo Zarzuela, and Yaroslav Tserkovnyak}
\affiliation{Department of Physics and Astronomy, University of California, Los Angeles, California 90095, USA}
%\date{\today}

\begin{abstract}
We analyze the electron dynamics in corrugated layers of transition-metal dichalcogenides. Due to the strong spin-orbit coupling, the intrinsic (Gaussian) curvature leads to an emergent gauge field associated with the Berry connection of the spinor wave function. We discuss the gauge field created by topological defects of the lattice, namely, tetragonal/octogonal disclinations and edge dislocations. Ripples and topological disorder induce the same dephasing effects as a random magnetic field, suppressing the weak localization effects. This geometric magnetic field can be detected in a Aharonov-Bohm interferometry experiment by measuring the local density of states in the vicinity of corrugations.
\end{abstract}
%\pacs{}

\title{Emergent gauge fields from curvature in \\single layers of transition-metal dichalcogenides}

\maketitle

\textit{Introduction}. Graphene and other two-dimensional crystals have served as new platforms for the controlled interplay between mechanical and electronic properties in material science.\cite{rev1} Mechanical distortions are usually incorporated as a background geometry in the effective theory describing the long-wavelength dynamics of electrons.\cite{deJuan_etal} In the particular case of graphene, gauge-like fields emerge due to the effect of mechanical tensions,\cite{tensions} corrugations,\cite{corrugations} or topological defects\cite{topo_defects} on the lattice. These fields arise as a manifestation of the chirality of the Bloch wave functions around the two inequivalent corners of the hexagonal Brillouin zone, $\mathbf{K}_{\pm}$. Single-layers of transition-metal dichalcogenides\cite{rev2} (TMDCs) combine these features with the strong spin-orbit coupling provided by the transition-metal atoms. The latter removes the spin degeneracy of the bands due to the lack of a center of inversion in the unit cell, while preserving the spin quantum number along the out-of-plane direction because of its mirror symmetry.\cite{Xiao_etal} This observation has been exploited in different proposals for spintronics and optoelectronics applications.\cite{applications} In particular, the application of tensions with trigonal symmetry leads to the formation of pseudo-Landau levels and the possibility of a quantum spin Hall effect,\cite{Cazalilla_etal} similarly to the original proposal in bulk zinc-blende semiconductors.\cite{Zhang}

Accumulation of Berry phases\cite{Berry} in real space yields an additional source of gauge fields. For instance, in corrugated graphene, where the spin-orbit interaction is negligible, these Berry phase effects are induced by the motion of the atomic orbital basis along with the distorted lattice.\cite{Trif_etal} This contribution, however, is purely dynamical and parametrically small at low frequencies in comparison to the aforementioned pseudo-gauge fields. In the case of TMDCs the situation may reverse: first of all, the chirality is not longer a good quantum number, so its effects are strongly attenuated for carriers close to the band edges due to the sizable gap in the band structure; second and more importantly, in the presence of corrugations the large spin-orbit coupling rotates the wave function in the spinor basis, which may engender a gauge field. We demonstrate that this is the case in this Letter: starting from a minimal Hamiltonian for the hexagonal (2H phase) crystal that incorporates the effect of curvature in the dynamics of low-energy excitations, we derive the emergence of a U(1) gauge field related to the spin-Berry connection of the wave function. The associated magnetic field is intrinsically determined by the geometry of the distorted lattice. Experimental consequences are discussed with emphasis on topological defects, which have been observed by transmission electron microscopy.\cite{dislocations} Our theory falls within a broader framework of interplay between topology and orientational (spin or nematic) order, with examples in nano-magnetism,\cite{mobius} liquid crystals,\cite{deGennes} or more recently in Weyl semimetals.\cite{Hughes}

\begin{figure}[t!]
\begin{center}
\hspace{-0.4cm}
\includegraphics[width=9cm]{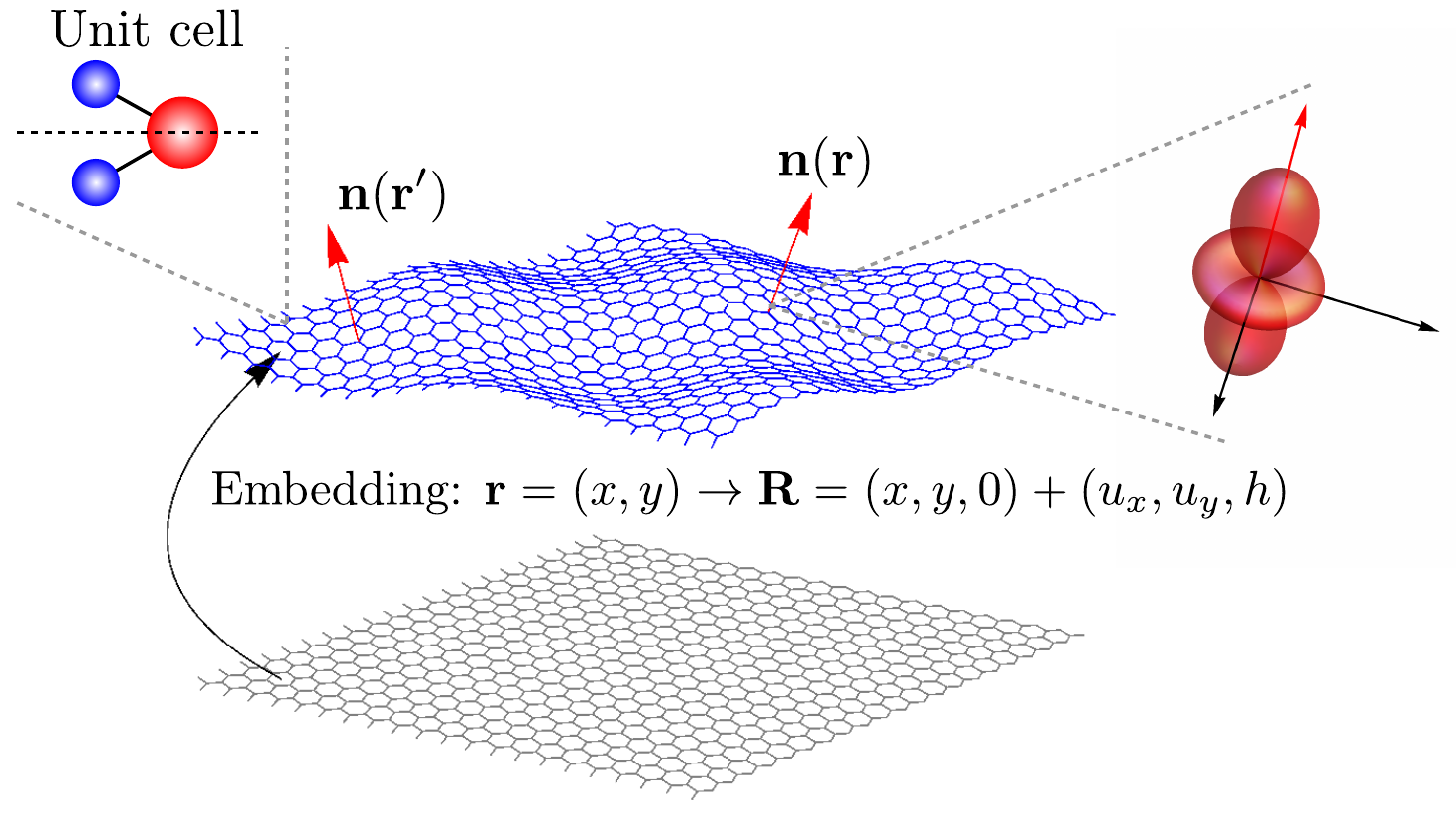}
\caption{Corrugated TMDC crystal in the 2H phase described as a membrane in the Monge's representation. Red arrows represent normal vectors to the surface. The unit cell consists of one transition-metal and two chalcogen atoms. The dashed line represents the plane of mirror reflection. The lowest energy bands are dominated by $d$ orbitals from the transition-metal atoms. The local orbital basis (in the figure, a $d_{z^2}$ orbital dominating the conduction band) is oriented along the normals.}
\vspace{-0.5cm} 
\label{Fig1}
\end{center}
\end{figure}

\textit{Model}. We are interested in the effective low-energy description of electrons in a corrugated TMDC with account of spin-orbit interactions. The $\mathbf{k}\cdot\mathbf{p}$ Hamiltonian around the two inequivalent valleys reads\cite{Xiao_etal,Ochoa_etal}
\begin{equation}
\label{eq1}
\hat{\mathcal{H}}=\frac{\hbar^{2}\bm{k}^{2}}{2m^{*}}\pm\Delta_{\textrm{so}}\hspace{0.05cm}\hat{\bm{s}}\cdot \bm{n}(\bm{r}),
\end{equation}
where $\bm{k}=-i\boldsymbol{\partial}$ is the momentum operator around the $\mathbf{K}_{\pm}$ points and $m^{*}$ is the effective mass of carriers near the edges of the lowest energy bands. The last term accounts for the spin-orbit coupling, where $\hat{\bm{s}}$ is the vector of Pauli matrices associated with the spin degree of freedom, the $\pm$ sign applies to $\mathbf{K}_{\pm}$ valleys, and $\bm{n}\left(\bm{r}\right)$ represents the normal to the crystal surface. Here $\bm{r}=(x,y)$ are the positions of the unit cells in the crystalline configuration, which parametrizes the surface in the so-called Monge's representation, see Fig.~\ref{Fig1}.

The Hamiltonian in Eq.~\eqref{eq1} is expressed in an internal frame of reference, in which the local orbital basis is rotated with respect to an inertial (laboratory) frame, as depicted in Fig.~\ref{Fig1}. In the pristine crystalline phase, $\bm{n}\left(\bm{r}\right)\equiv\bm{e}_z$, the Hamiltonian is compatible with the $D_{3h}=D_3\times\sigma_h$ point group symmetry of the lattice. Real samples may however present corrugations due to either the interaction with a substrate or thermal fluctuations, breaking the $\sigma_h$ (mirror $z\rightarrow-z$) symmetry. As bands of opposite parity with respect to mirror reflection are well separated in energy, orbital hybridization can be safely neglected.\cite{SM} The mirror-symmetry breaking is then incorporated by locking the quantization axis to the normal, as imposed by the spin-orbit interaction. A similar approach has been taken in carbon nanotubes.\cite{Rudner_Rashba} In this Letter we focus on the static case, so that $\bm{n}\left(\bm{r}\right)$ depends only on spatial coordinates. 

\textit{Adiabatic limit}. It is convenient to study the model in the spinor basis adjusted to the local quantization axis defined by $\bm{n}\left(\bm{r}\right)$. For this purpose, let us consider a local unitary rotation satisfying\cite{unitary}
\begin{align}
\label{eq2}
\hat{\mathcal{U}}^{\dagger}(\bm{r})[\hat{\bm{s}}\cdot \bm{n}(\bm{r})]\hat{\mathcal{U}}(\bm{r})=\hat{s}_{z}.
\end{align}
The unitary operator $\hat{\mathcal{U}}\left(\bm{r}\right)$ implements a SU(2)/U(1) gauge transformation on the spinor wave function. The transformed Hamiltonian is
\begin{equation}
\label{eq3}
\hat{\mathcal{U}}^{\dagger}(\bm{r})\hat{\mathcal{H}}\,\hat{\mathcal{U}}(\bm{r})=\frac{\hbar^{2}}{2m^{*}}\big[\bm{k}-\hat{\bm{A}}(\bm{r})\big]^{2}\pm\Delta_{\textrm{so}}\hat{s}_{z},
\end{equation}
where the components of the gauge field read
\begin{equation}
\label{eq4}
\hat{A}_{\mu}(\bm{r})\equiv i\,\hat{\mathcal{U}}^{\dagger}(\bm{r})\partial_{\mu}\,\hat{\mathcal{U}}(\bm{r})=\hat{A}_{\mu}^{\textrm{aff}}\left(\bm{r}\right)+A_{\mu}^{\textrm{B}}\left(\bm{r}\right)\hat{s}_z.
\end{equation}
We emphasize that $\hat{A}_{\mu}(\bm{r})$ splits into two different contributions: $\hat{A}^{\textrm{aff}}_{\mu}(\bm{r})$ is univocally determined by the extrinsic curvature of the surface,\cite{book} and therefore does not depend on the particular choice of $\hat{\mathcal{U}}\left(\bm{r}\right)$; on the contrary, $A_{\mu}^{\textrm{B}}\left(\bm{r}\right)$ does depend explicitly on this choice, which manifests the U(1) ambiguity of the theory.\cite{U1}

The gauge-independent (transverse) contribution comes from the affine connection defined by the corrugated surface. It accounts for the deviation of the spin polarization vector with respect to $\bm{n}$. This field can be recast in terms of the extrinsic curvature tensor (second fundamental form) as\cite{SM}
\begin{align}
\label{eq6}
\hat{A}_{\mu}^{\textrm{aff}}(\bm{r})&=\frac{i}{4}f_{\mu\nu}\left(\bm{r}\right)[\hat{s}_z,\hat{s}_{\nu}]\approx\frac{i}{4}\partial_{\mu}\partial_{\nu}h\left(\bm{r}\right)[\hat{s}_z,\hat{s}_{\nu}].
\end{align}
The last expression comes from an expansion to the lowest order in the displacement fields in the Monge's representation, where the brackets $[\,,\,]$ denote the commutator of Pauli matrices.

This deviation induces spin flip processes through a coupling of the form $\hat{\bm{A}}^{\textrm{aff}}(\bm{r})\cdot\bm{k}$. Due to the large spin-orbit splitting, however, spin relaxation can be neglected in the limit of smooth corrugations defined by
\begin{equation}
\label{eq6}
\nabla^2h\ll\frac{\Delta_{\textrm{so}}}{\hbar v_F},
\end{equation}
where $v_F$ is the Fermi velocity of carriers. In this regime, the spin follows adiabatically the local quantization axis imposed by the spin-orbit interaction. The wave function acquires a phase while moving over a curved section of the distorted crystal, which is associated with the spin-Berry connection provided by the remaining U(1) gauge contribution in Eq.~\eqref{eq4}. As a result, a perpendicular magnetic field arises in the dynamics of carriers with respect to the co-moving internal frame,\cite{SM}
\begin{equation}
\label{eq7}
\mathcal{B}(\bm{r})=\pm\frac{\hbar}{e}\,\nabla\times\bm{A}^{\textrm{B}}(\bm{r})=\pm\frac{\hbar}{2|e|}\kappa(\bm{r}).
\end{equation}
Here $\kappa(\bm{r})$ is the Gaussian curvature and the sign $+(-)$ corresponds to spin up (down) electrons with respect to the local quantization axis. Notice that this is an intrinsic geometrical property of the distorted crystal.

\begin{figure}[t!]
\begin{center}
\hspace{-0.4cm}
\includegraphics[width=9cm]{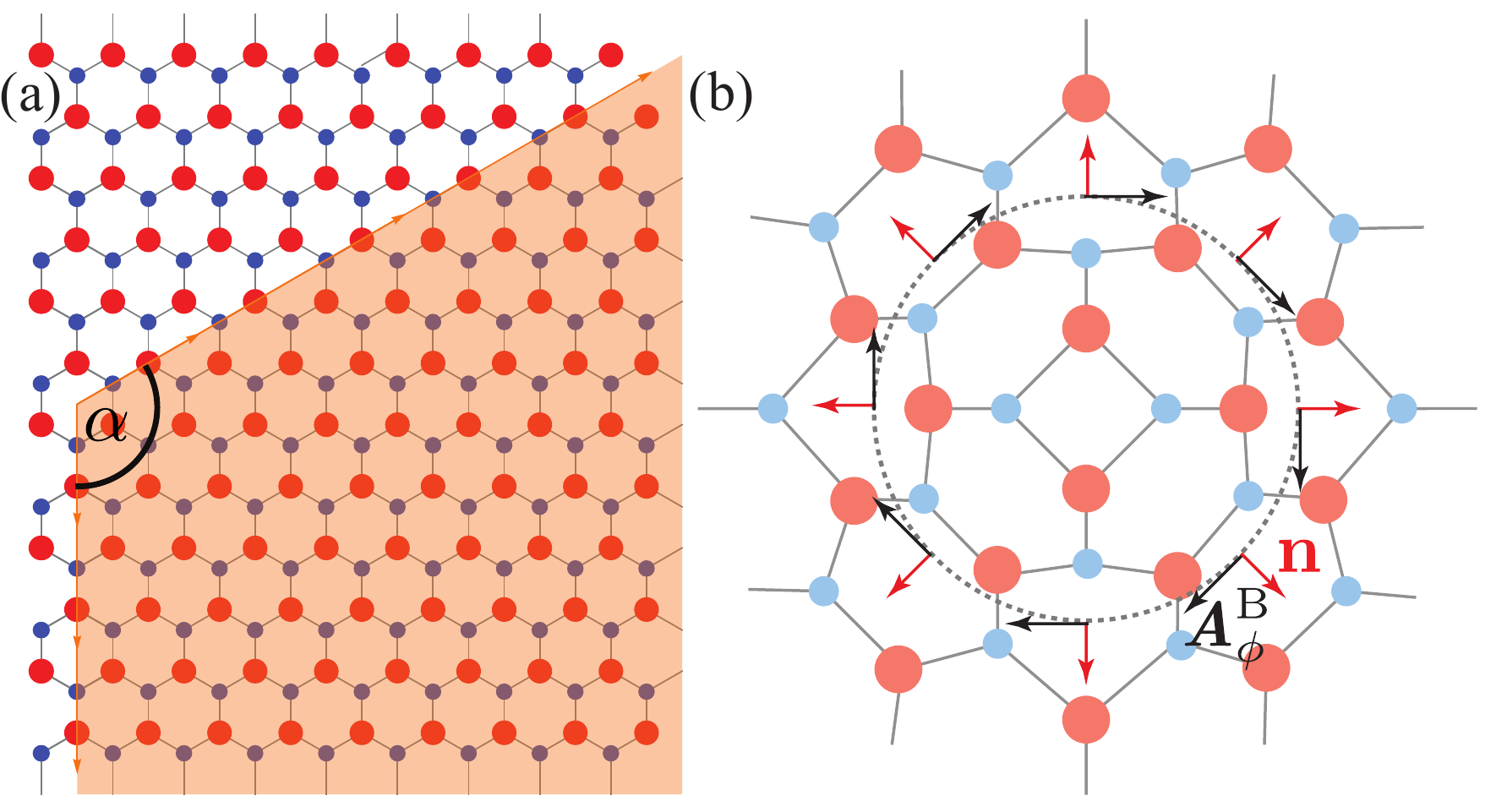}
\caption{(a) Tetragonal disclination in the Volterra's construction: a sector of the lattice with central angle $\alpha=\frac{2\pi}{3}$ is removed and then the boundaries are identified. (b) Top view of the resulting cone-like shape. The red arrows represent the normal to the surface, and the black ones point along the vortex field, $A_{\phi}^{\textrm{B}}=-\frac{\alpha}{4\pi\left|\bm{r}\right|}$.}
\vspace{-0.6cm} 
\label{Fig2}
\end{center}
\end{figure}

\begin{figure}[t!]
\begin{center}
%\vspace{1cm}
\includegraphics[width=8.5cm]{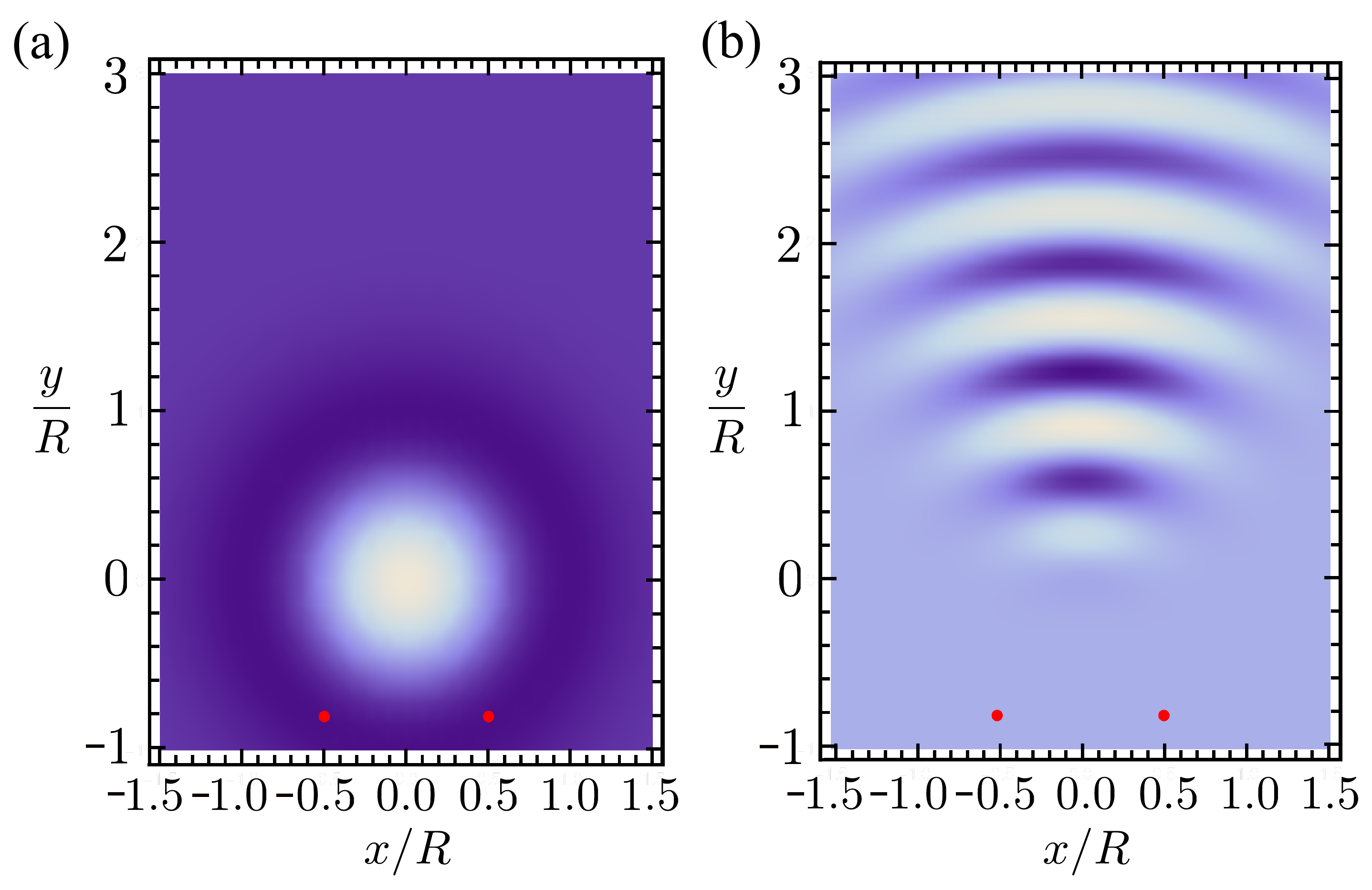}
%\vspace{4.1cm} 
\caption{%(a) Aharonov-Bohm interferometer proposed in the text. {\blue The blue and red spheres represent the scatterers}.
(a) Magnetic field profile due to a gaussian bump, described by a heigh t profile of the form $h\left(\bm{r}\right)=h_0e^{-|\bm{r}|^2/R^2}$. (b) Map of the local density of states at the Fermi level in arbitrary units. The model parameters are $h_0/R=0.1$, $k_{F}R=5$, $m^{*}U_0/2\pi\hbar^2=1$. The red dots represent the positions of the scattering centers.}
\vspace{-0.5cm} 
\label{Fig3}
\end{center}
\end{figure} 

\textit{Topological defects}. The magnetic flux across the surface of the distorted crystal is quantized according to the Gauss-Bonnet theorem,\cite{SM}
\begin{equation}
\label{eq8}
\Phi=\int_{S}\mathcal{B}(\bm{r})\ud A=\frac{\hbar}{2|e|}\int_{S}\kappa(\bm{r})\ud A=\chi(S)\Phi_{0}.
\end{equation}
Here $\chi(S)$ is the Euler characteristic of the corrugated surface, $\Phi_{0}=h/2|e|$ is the flux quantum, and we have neglected contributions from the boundary. Any configuration smoothly connected to the flat phase has zero Euler characteristic and thus the total flux is zero. %Hence, the onset of Hall effect physics requires the introduction of defects into the lattice that modify the topology of the surface.
Topological defects --disclinations and dislocations-- are therefore a natural source of curvature. Isolated disclinations can be introduced into the crystal lattice via the Volterra's (cut-and-paste) construction: as depicted in Fig. \ref{Fig2}~(a), these defects can be built by removing (adding) a sector of the lattice with central angle $\alpha$ --also referred to as the defect (excess) angle-- and then by identifying its boundaries. This leads to a cone-like shape, see Fig.~\ref{Fig2}~(b), whose Gaussian curvature is zero everywhere except at the cusp, $\kappa(\bm{r})=\alpha\,\delta^{(2)}(\bm{r})$.\cite{SM} Hence, we identify $\alpha$ with the topological charge of the defect.

In TMDC crystals, defects preserving the coordination of the lattice are the tetragonal ($\alpha_{4}=\frac{2\pi}{3}$) and octogonal ($\alpha_{8}=-\frac{2\pi}{3}$) disclinations. The competition between stretching and bending energies relaxes the position of the atoms. The involved strain fields are long-ranged, similarly to the case of a screw dislocation in a 3D solid.\cite{Chaikin_Lubensky} The long-ranged nature of the defect is also present in the Berry-connection gauge field, which possesses the structure of a vortex, see Fig.~\ref{Fig2}~(b). The topological charge of the disclination can be understood on physical grounds as twice the Aharonov-Bohm phase acquired by the wave function when the electron surrounds the defect, reminiscent of the removed angular sector in the Volterra's construction. With account of Eq. \eqref{eq8} the total spin-Berry magnetic flux reads 
\begin{equation}
\label{eq14}
\Phi=\frac{1}{6}(n_{4}-n_{8})\Phi_{0}^{\textrm{Dirac}},
\end{equation}
where $\Phi_{0}^{\textrm{Dirac}}=h/|e|$ is the Dirac monopole flux quantum and $n_{4}$ ($n_{8}$) is the number of tetragonal (octogonal) disclinations. In asymptotically flat samples we have $n_4=n_8$, and therefore the total flux is zero. A pair of complementary disclinations form an edge dislocation with the Burgers vector perpendicular to the axis linking the centers of the pair. On the contrary, an imbalance in the topological charge bends the TMDC layer, eventually folding the membrane into a closed molecule.
%We consider two scenarios: \textit{i)} Asymptotically flat samples. This requires the same number of tetragonal and octogonal defects, and therefore the total flux is zero. \textit{ii)} Folding of the TMDC layer into a closed molecule due to the nucleation of defects.
 This happens when the condition $n_{4}-n_{8}=6$ is met according to Euler's formula.\cite{SM} Through Eq. \eqref{eq14} we conclude that the topological closure is equivalent to the Dirac quantization condition, i. e. electrons experience the magnetic field created by a Dirac monopole located at the center of the molecule. %Hence, the emergent magnetic field corresponds to the one produced by a Dirac monopole located at the center of the molecule.

\begin{figure*}[t!]
\begin{center}
%\vspace{1cm}
\includegraphics[width=18cm]{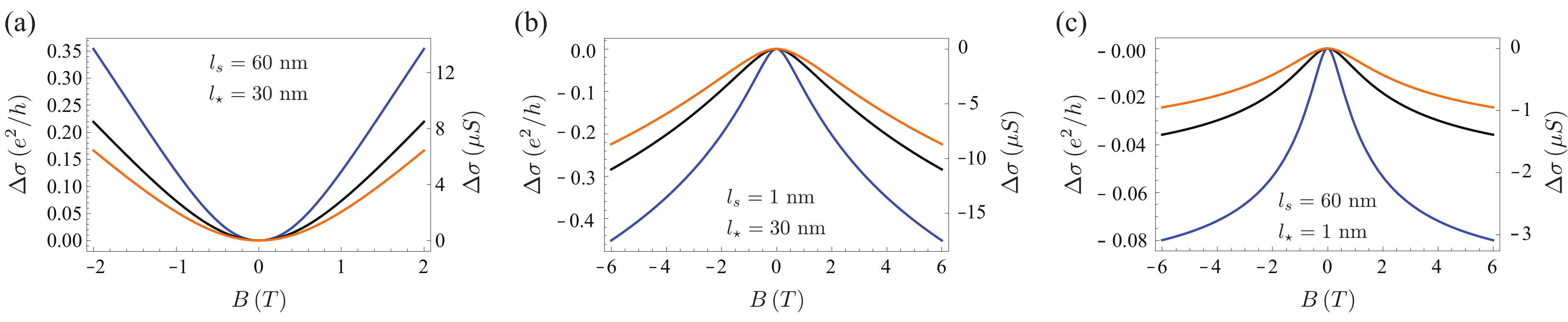}
%\vspace{4.1cm} 
\caption{Magneto-conductance of single-layer TMDCs at $T = 1.5$ K (blue), $T = 3$ K (black) and $T = 4$ K (orange) calculated via Eq~\eqref{eq:delta_sigma} for the 3 regimes discussed in the text. In all cases we have considered $l_{\varphi}[\textrm{nm}] = 50/\sqrt{2.5\times T [\textrm{K}]}$ in agreement with magneto-transport experiments in multi-layers.\cite{few}}
\vspace{-0.5cm} 
\label{Fig4}
\end{center}
\end{figure*}

%{\red Note: Maybe here we could discuss the spin pumping driven by the motion of topological defects ...}

%{\red Maybe a footnote with the spin Hall...}

\textit{Aharonov-Bohm effect}. Electrons and holes surrounding corrugated regions of the sample acquire a non-trivial phase %even in the case of a topologically trivial profile, which 
that gives rise to quantum interference corrections to spectral properties. We propose a scanning tunnel microscopy (STM) experiment to probe the Aharonov-Bohm effect\cite{AB} on the local density of states induced by the proximity of a corrugation. Our scheme follows closely previous proposals to measure the same effect due to real\cite{Cano} and pseudo-magnetic fields in graphene.\cite{Cano_etal}
%Generation of a quantum interference pattern requires the presence of scattering centers along the sample.
The simplest Aharonov-Bohm interferometer consists of two scattering centers. Processes corresponding to semiclassical closed paths between the scattering centers and the STM tip give rise to a correction in the local density of states due to the presence of curvature, $\delta\rho=\rho-\rho_{\textrm{flat}}$, which is evaluated as
\begin{equation*}
%\label{eq15}
\delta\rho\left(\bm{r},\omega\right)\sim-\frac{1}{\pi}\text{ Im }\delta G_{\textrm{loop}} \left(\bm{r},\bm{r},\omega\right)\times\left[\cos\frac{\pi\Phi_{\textrm{loop}}}{\Phi_0}-1\right].
\end{equation*}
Here $\delta G_{\textrm{loop}} \left(\bm{r},\bm{r},\omega\right)$ represents the correction to the local Green function due to these scattering loops, which can be computed by means of standard diagrammatic techniques.\cite{SM} In the derivation of this expression we employed the semiclassical approximation for the Green function,
%\begin{align*}
%G\left(\bm{r}_1,\bm{r}_2,\omega\right)=\text{Exp}\left[i\int_{\bm{r}_1}^{\bm{r}_2}d\bm{l}\cdot\bm{A}^{\textrm{B}}(\bm{r})\right]G_0\left(\bm{r}_1,\bm{r}_2,\omega\right),
%\end{align*}
which is justified in the adiabatic regime, Eq.~\eqref{eq6}. The cosine factor stems from the fact that trajectories enclosing anti-clockwisely the corrugation acquire a phase proportional to the magnetic flux across the enclosed area, $\Phi_{\textrm{loop}}$, whereas this phase has the opposite sign for clockwise trajectories.

We consider the case of a gaussian bump. Fig.~\ref{Fig3}~(a) shows the profile of the emergent magnetic field provided by Eq.~\eqref{eq7}. The correction to the local density of states is shown in Fig.~\ref{Fig3}~(b). The red dots represent the positions of the scattering centers. In the calculation we have assumed a scattering potential of the form $U_0\,\delta\left(\bm{r}-\bm{r}_i\right)$ and we have neglected inter-valley scattering, which is justified close to the bottom of the band.\cite{Koshino} Notice that the effect is not sensitive to the sign of the flux, so contributions from carriers with opposite spin sum up.

\textit{Magneto-transport}. The apparent time-reversal symmetry breaking induced by corrugations and topological disorder suppresses the longitudinal magneto-conductance, $\Delta \sigma_{xx}\left(B\right)\equiv \sigma_{xx}\left(B\right)-\sigma_{xx}\left(0\right)$, in the same manner as the scattering off magnetic impurities. Related mechanisms have been previously discussed in the context of multi-valley metals.\cite{valley} The low-field magneto-conductance arises due to the partial suppression of the weak (anti-)localization correction to the conductivity. The later results from the quantum interference on self-intersecting time-reversed diffusive trajectories, along which the phase coherence of the wave function is preserved, $l_{\varphi}\gg \ell$. Here $\ell$ represents the mean free path, characterizing the typical size of diffusive loops, and $l_{\varphi}$ is the phase coherence length limited by inelastic scattering off phonons or due to electron-electron interactions. %The latter seems to be the dominant mechanism in TMDCs according to magneto-transport experiments in multi-layer MoS$_2$.\cite{few} 
The fictitious time-reversal symmetry breaking caused by the Gaussian curvature of the sample is only effective when the spin is conserved, $l_s\gg\ell$. Here $l_s$ represents the spin diffusion length accounting for both inter- and intra-valley spin relaxation mechanisms. The random magnetic fields associated with the curvature of the sample define another length scale, the dephasing length $l_{\star}$. %, which can be estimated as\begin{align*}
%l_{\star}\sim\begin{cases}
%\frac{k_F R^6}{h_0^4}&\text{in the case of ripples},\\
%k_F d^2&\text{in the case of dislocations},
%\end{cases}
%\end{align*}
%$k_F$ being the Fermi wave vector.
This irreversible dephasing mechanism opens relaxation gaps in the spin/valley mixing triplet channels of the Cooperon correlation function, leading to the magneto-conductance formula\cite{SM,WL}
\begin{gather}
 \Delta \sigma_{xx}\left(B\right)=\frac{e^2}{\pi h}\left[4F\left(\frac{B}{B_{\varphi}+B_{s}+B_{\star}}\right)
\right.\nonumber\\\left.
 +F\left(\frac{B}{B_{\varphi}+2B_{s}}\right)-F\left(\frac{B}{B_{\varphi}}\right)
\right],
\label{eq:delta_sigma}
\end{gather}
where $F\left(x\right)\equiv\ln\left(x\right)+\psi\left(\frac{1}{2}+\frac{1}{x}\right)$, $\psi\left(x\right)$ is the digamma function, and $B_{\alpha}\equiv\hbar/4el_{\alpha}^2$.

Fig.~\ref{Fig4} shows the magneto-conductance deduced from Eq.~\eqref{eq:delta_sigma} for the 3 possible regimes: (a) $l_{\varphi}\ll l_{s},l_{\star}$, where the spin quantum number and phase coherence of the wave function are preserved along the diffusive loops responsible for the weak localization%and therefore positive magneto-conductance
. (b) $l_{s}\ll l_{\varphi},l_{\star}$, where the apparent time-reversal symmetry breaking is ineffective due to spin relaxation, which engenders a phase difference $\Delta\varphi\sim\pi$ between time-reversed trajectories and therefore a destructive interference, leading to weak anti-localization and negative magneto-conductance at low fields. (c) $l_{\star}\ll l_{s},l_{\varphi}$, where the electrons accumulate random phases within phase coherent trajectories due to the presence of curvature, suppressing the weak (anti-)localization as time-reversal symmetry for spin up and down electrons is effectively broken. In the case of MoS$_2$, multi-layer samples show the usual weak localization behavior\cite{few} characteristic of regime (a), whereas the magneto-conductance of the single-layer is 1-2 orders of magnitude lower,\cite{single1,single2} compatible with regime (c). Interestingly, the negative magneto-conductance in single-layer MoS$_2$ reported in Ref.~\onlinecite{single2} is reproduced if we take the same inelastic coherence and spin diffusion lengths as in the multi-layer, see panel Fig.~\ref{Fig4}~(c).

\textit{Discussion}. TMDC samples show similar corrugations as in graphene, with average heights of the order of $h_0\sim1$ nm and lateral sizes $R~\sim 2-20$ nm.\cite{ripples} The effective magnetic field at the center of these ripples is in the ballpark of $\mathcal{B}\sim\Phi_0\left(\frac{h_0}{R^2}\right)^2\sim0.1-1$ T. Even larger fields are within experimental reach by applying compression. The large spin-orbit splittings --tens of meV in the conduction band and hundreds in the valence band-- ensure the adiabatic condition for carriers close to the band edges. Interestingly, the modulation of the Fermi level leads to a crossover from the adiabatic regime, in which the gauge field stems from Berry phases in real space, to a regime dominated by pseudo-gauge fields caused by strain. This crossover should be detected in the proposed STM experiment as a variation in the interference pattern on account of the lower symmetry of the strain-induced pseudo-magnetic field, reminiscence of the structure of the wave function in reciprocal space.
%In the case of cylindrically symmetric corrugations, the strain-induced pseudo-magnetic field possesses only trigonal symmetry. This is a reminiscence of the structure of the wave function in reciprocal space. Therefore, the crossover should be detected in the proposed STM experiment as a variation in the Aharonov-Bohm interference pattern.

We have focused on the geometric (static) aspects of the gauge theory in this Letter. Spin-dependent electromotive forces generated by dynamical membranes will be object of future research. For example, the dynamics of topological defects should pump spin analogously to the motion of solitons in magnets.\cite{Brataas} Reciprocally, spin currents will generate a back-action in the membrane dynamics, leading to melting instabilities due to the proliferation of topological defects. Finally, an imbalance in the number of disclinations with opposite charge will give rise to a spin-Hall response.\cite{spin-Hall} %As a final remark, topological defects are likely to be present in grain boundaries between different structural phases of these materials.\cite{acs}

In summary, we have shown the emergence of a gauge field associated with the spin-Berry connection in corrugated TMDC crystals. The phases accumulated by electrons and holes moving along curved sections of the crystal give rise to corrections in their single- and two-particle properties due to quantum interference. The former can be detected as a variation in the local density of states in the vicinity of corrugations and topological defects, whereas the latter may explain the suppressed localization effects observed in magneto-transport experiments in single-layer MoS$_2$,\cite{single1,single2} in contrast to the conventional weak localization behavior in the multi-layer counterpart.\cite{few} This analysis could be extended to other systems such as zinc-blende semiconductors and surface states of topological insulators. %Effects beyond the static regime such as the electromotive forces induced by dynamical corrugations, and reciprocally, the back-action of spin currents on the dynamics of topological defects, will be object of future research.

\textit{Acknowledgements}. This work was supported by the U.S. Department of Energy, Office of Basic Energy Sciences under Award No. DE-SC0012190 (H.O.), and by the NSF-funded MRSEC under Grant No. DMR-1420451 (R.Z.). R.Z. also thanks Fundaci\'{o}n Ram\'{o}n Areces for a postdoctoral fellowship within the XXVII Convocatoria de Becas para Ampliaci\'{o}n de Estudios en el Extranjero en Ciencias de la Vida y de la Materia.

\clearpage

\section*{Supplementary material}

\subsection{Elements of geometry}

In a continuum description, the unit cells of the 2D crystal are assumed to occupy positions in a smooth surface $S$ defined by the embedding $\bm{R}\left(\bm{r}\right)$. Here $\bm{r}$ must be taken as the intrinsic coordinates parameterizing the surface. In the Monge's representation, this corresponds to the positions of the unit cell in the crystalline configuration.

Any smooth surface is locally flat, i.e. at each point $\bm{r}$ it can be identified with its tangent plane, $T_{\bm{r}}S$. This plane is spanned by the vectors $\partial_i\bm{R}\left(\bm{r}\right)$, which is referred to as the intrinsic basis. An element $\bm{u}$ of $T_{\bm{r}}S$ can be decomposed as $\bm{u}=u_i\partial_i\bm{R}$, where repeated indices are summed up. In our notation, $u_i$ arranged as a column vector is denoted by $\vec{u}$.

\subsubsection{Fundamental forms}

Next, we introduce the two fundamental forms of the surface, i.e. bi-linear applications defined over the tangent space that define the geometrical properties of the surface as an embedded manifold.

\subsubsection*{First fundamental form}

The first fundamental form,\begin{align}
I\left(\vec{u},\vec{v}\right)=\bm{u}\cdot\bm{v}=\vec{u}^T\cdot\hat{g}\left(\bm{r}\right)\cdot\vec{v},
\end{align}
is the pull-backed metric tensor of the Euclidian space onto the surface. Its intrinsic components read\begin{align}
\label{eq:g}
g_{ij}\left(\bm{r}\right)=\partial_i\bm{R}\left(\bm{r}\right)\cdot\partial_j\bm{R}\left(\bm{r}\right).
\end{align}
This form establishes how to measure lengths and angles on the surface and therefore defines the intrinsic geometrical properties of the embedding. In the Monge's representation, the first fundamental form is related to the strain tensor via the identity\begin{align}
g_{ij}\left(\bm{r}\right)=\delta_{ij}+2u_{ij}\left(\bm{\bm{r}}\right),
\end{align}
where the components of the strain tensor are defined as\begin{gather}
u_{ij}\left(\bm{r}\right)=\frac{1}{2}\left(\partial_iu_j\left(\bm{r}\right)+\partial_ju_i\left(\bm{r}\right)+\partial_ih\left(\bm{r}\right)\partial_jh\left(\bm{r}\right)\right)
\nonumber\\
+\,\,\partial_iu_k\left(\bm{r}\right)\partial_ju_k\left(\bm{r}\right).
\end{gather}
In the theory of elasticity, the last term is usually dropped.

\subsubsection*{Second fundamental form}

The second fundamental form is defined as\begin{align}
\label{eq:II}
II\left(\vec{u},\vec{v}\right)=-d_{\bm{u}}\bm{n}\left(\bm{r}\right)\cdot\bm{v}=\vec{u}^T\cdot\hat{f}\left(\bm{r}\right)\cdot\vec{v},
\end{align}
where $d_{\bm{u}}\bm{n}\left(\bm{r}\right)$ denotes the differential of the Gauss map,
\begin{align}
\bm{n}\left(\bm{r}\right)=\epsilon_{ij}\frac{\partial_i\bm{R}\left(\bm{r}\right)\times\partial_j\bm{R}\left(\bm{r}\right)}{\left|\epsilon_{ij}\partial_i\bm{R}\left(\bm{r}\right)\times\partial_j\bm{R}\left(\bm{r}\right)\right|},
\end{align}
along the vector $\bm{u}$. It measures the variations of the intrinsic vectors $\partial_i\bm{R}$ along the surface, projected onto the normal $\bm{n}\left(\bm{r}\right)$,\begin{align}
f_{ij}\left(\bm{r}\right)=\bm{n}\left(\bm{r}\right)\cdot\partial_i\partial_j\bm{R}\left(\bm{r}\right).
\end{align}

The second fundamental form can also be introduced as the vector field of tensors $\bm{f}_{ij}\left(\bm{r}\right)=f_{ij}\left(\bm{r}\right)\bm{n}\left(\bm{r}\right)$ through the Gauss-Codazzi equations, 
\begin{align}
\label{eq:f_field}
\bm{f}_{ij}\left(\bm{r}\right)=\partial_i\partial_j\bm{R}\left(\bm{r}\right)-\Gamma_{ijk}\left(\bm{r}\right)\partial_k\bm{R}\left(\bm{r}\right),
\end{align}
where $\Gamma_{ijk}\left(\bm{r}\right)$ are the Christoffel symbols of the Levi-Civita connection defined by the first fundamental form,\begin{align*}
\Gamma_{ijk}\left(\bm{r}\right)=\frac{1}{2}g_{kl}^{-1}\left(\bm{r}\right)\left(\partial_ig_{jl}\left(\bm{r}\right)+\partial_jg_{il}\left(\bm{r}\right)-\partial_lg_{ij}\left(\bm{r}\right)\right).
\end{align*}
From Eq.~\eqref{eq:f_field} we see that the second fundamental form is a measure of the extrinsic curvature of the surface.

\subsubsection{Weingarten map}

In the definition of the second fundamental form, notice that both $\bm{u}$ and $d_{\bm{u}}\bm{n}\left(\bm{r}\right)$ belong to the tangent space. The differential of the Gauss map defines the so-called Weingarten map,
\begin{align}
d_{\bm{u}}\bm{n}\left(\bm{r}\right)=-\hat{S}\left(\bm{r}\right)\cdot\vec{u},
\end{align}
where $\hat{S}\left(\bm{r}\right)$ is the shape operator. In components, this last equation reads\begin{align}
\label{eq:W1}
\partial_{i}\bm{n}\left(\bm{r}\right)=-S_{ji}\left(\bm{r}\right)\partial_j\bm{R}\left(\bm{r}\right).
\end{align}
The shape operator is related to both the first and second fundamental forms -- see Eq.~\eqref{eq:II} -- through the Weingarten equations,\begin{align}
\label{eq:W2}
\hat{S}\left(\bm{r}\right)=\hat{g}^{-1}\left(\bm{r}\right)\cdot\hat{f}\left(\bm{r}\right).
\end{align}

In Monge's representation to the lowest order in displacements, we have\begin{align}
\label{eq:expansion}
f_{ij}\left(\bm{r}\right)\approx S_{ij}\left(\bm{r}\right)\approx\partial_i\partial_j h\left(\bm{r}\right).
\end{align}

\subsubsection{Gaussian curvature}

The Gaussian curvature is defined as the determinant of the Shape operator:
\begin{align}
\kappa\left(\bm{r}\right)=\text{det }\hat{S}\left(\bm{r}\right).
\end{align}
Notice that this is an invariant under change of coordinates, as it is deduced from Eq.~\eqref{eq:W2}. Furthermore, Gauss's \textit{Theorema Egregium} establishes that $\kappa\left(\bm{r}\right)$ is an intrinsic invariant of the surface.

\subsubsection{Internal frame}

The Hamiltonian in Eq.~(1) of the main text is expressed in an internal orbital basis. Therefore, the dynamics of the electron are observed from an internal, co-moving frame of reference. The crystalline momentum operator is defined as $\bm{k}=-i\bm{e}_{\mu}\left(\bm{r}\right)\partial_{\mu}$, where $\bm{e}_{\mu}\left(\bm{r}\right)$ are unit vectors defining the internal frame at each point (known as the frame fields), and $\partial_{\mu}$ are derivatives with respect to the internal coordinates, locally defined through the transformation $\zeta^{\mu}\left(\bm{r}\right)$. The unit vectors $\bm{e}_{\mu}\left(\bm{r}\right)$ form an orthonormal basis of $T_{\bm{r}}S$, in such a way that
\begin{align}
\label{eq:frame}
\partial_i\bm{R}\left(\bm{r}\right)=\mathfrak{e}_i^{\mu}\left(\bm{r}\right)\bm{e}_{\mu}\left(\bm{r}\right),
\end{align}
where $\mathfrak{e}_i^{\mu}\left(\bm{r}\right)=\partial_i \zeta^{\mu}\left(\bm{r}\right)$ is the Jacobian of the coordinate transformation.

The choice of the internal frame is not univocal, but must be consistent with the gauge choice in the unitary transformation of the Hamiltonian. This implies that\begin{align}
\label{eq:Re}
\bm{e}_{\mu}\left(\bm{r}\right)=\hat{\mathcal{R}}\left(\bm{r}\right)\boldsymbol{\mu},\qquad \mu=x,y,
\end{align}
where $\hat{\mathcal{R}}\left(\bm{r}\right)$ is the SO(3) matrix associated with the unitary transformation, defined by the relation\begin{align}
\label{eq:R}
\bm{n}\left(\bm{r}\right)=\hat{\mathcal{R}}\left(\bm{r}\right){\bm{z}}.
\end{align}

From Eqs.~\eqref{eq:g} and \eqref{eq:frame} we obtain the relation,\begin{align*}
g_{ij}\left(\bm{r}\right)=\mathfrak{e}_i^{\mu}\left(\bm{r}\right)\mathfrak{e}_j^{\nu}\left(\bm{r}\right)\bm{e}_{\mu}\left(\bm{r}\right)\cdot\bm{e}_{\nu}\left(\bm{r}\right)=\mathfrak{e}_i^{\mu}\left(\bm{r}\right)\mathfrak{e}_j^{\mu}\left(\bm{r}\right),
\end{align*}
or in matrix notation ($\left[\hat{\mathfrak{e}}\right]_{\mu i}\equiv\mathfrak{e}_i^{\mu}$),\begin{align}
\hat{g}\left(\bm{r}\right)=\hat{\mathfrak{e}}^T\left(\bm{r}\right)\cdot \hat{\mathfrak{e}}\left(\bm{r}\right).
\end{align}
This last relation allows us to write the internal area element as
\begin{align}
dA=d\zeta^x d\zeta^y=\text{det }\hat{\mathfrak{e}}\left(\bm{r}\right) \,d^2\bm{r}=\sqrt{\text{det }\hat{g}\left(\bm{r}\right)} \,d^2\bm{r}.
\end{align}

\subsection{Gauge field}

The gauge field associated with the unitary transformation of the Hamiltonian is defined as\begin{align}
\hat{A}_{\mu}\left(\bm{r}\right)=i\,\hat{\mathcal{U}}^{\dagger}\left(\bm{r}\right)\partial_{\mu}\,\hat{\mathcal{U}}\left(\bm{r}\right).
\label{eq:A_def}
\end{align}
So far, Eq.~(2) in the main text does not determine unambiguously the transformation. Indeed, the unitary operator $\hat{\mathcal{U}}\left(\bm{r}\right)$ implements a SU(2)/U(1) gauge transformation of the spinor wave function, meaning that the gauge field $\hat{A}_{\mu}\left(\bm{r}\right)$ takes values in the SU(2) algebra, but the transformed Hamiltonian only possesses a U(1) gauge symmetry.

By inverting Eq.~(2) of the main text and taking derivatives, we obtain\begin{align*}
\hat{\bm{s}}\cdot\partial_{\mu}\bm{n}\left(\bm{r}\right)=\partial_{\mu}\,\hat{\mathcal{U}}\left(\bm{r}\right)
\hat{s}_z\,\hat{\mathcal{U}}^{\dagger}\left(\bm{r}\right)+\hat{\mathcal{U}}\left(\bm{r}\right)\,\hat{s}_z\,\partial_{\mu}\,\hat{\mathcal{U}}^{\dagger}\left(\bm{r}\right),
\end{align*}
which can be rewritten, by using the unitary property of $\hat{\mathcal{U}}\left(\bm{r}\right)$, as\begin{align}
\hat{\bm{s}}\cdot\partial_{\mu}\bm{n}\left(\bm{r}\right)=i\left[\bm{n}\left(\bm{r}\right)\cdot\hat{\bm{s}},
\tilde{A}_{\mu}\left(\bm{r}\right)\right].
\label{eq:conmutador}
\end{align}
Here we have introduced the \textit{rotated} field $\tilde{A}_{\mu}\left(\bm{r}\right)\equiv\hat{\mathcal{U}}\left(\bm{r}\right)\hat{A}_{\mu}\left(\bm{r}\right)\hat{\mathcal{U}}^{\dagger}\left(\bm{r}\right)$. It is convenient to introduce the following vectorial notation:\begin{align*}
\hat{A}_i\left(\bm{r}\right)=\bm{\hat{A}}_{\mu}\left(\bm{r}\right)\cdot\bm{s},\\
\tilde{A}_{\mu}\left(\bm{r}\right)=\bm{\tilde{A}}_i\left(\bm{r}\right)\cdot\bm{s}.
\end{align*}
Notice that the vectors $\bm{\hat{A}}_{\mu}\left(\bm{r}\right)$, $\bm{\tilde{A}}_{\mu}\left(\bm{r}\right)$, are related by a rotation of the form\begin{align*}
\bm{\tilde{A}}_i\left(\bm{r}\right)=\hat{\mathcal{R}}\left(\bm{r}\right)\bm{\hat{A}}_i\left(\bm{r}\right),
\end{align*}
where $\hat{\mathcal{R}}\left(\bm{r}\right)$ is defined in Eq.~\eqref{eq:R}.
From Eq.~\eqref{eq:conmutador} and with account of the algebra of commutators of Pauli matrices we can write\begin{align*}
\partial_{\mu}\bm{n}\left(\bm{r}\right)=-2\,\bm{n}\left(\bm{r}\right)\times\bm{\tilde{A}}_{\mu}\left(\bm{r}\right).
\end{align*}
By inverting this last relation we have\begin{align*}
\bm{\tilde{A}}_{\mu}\left(\bm{r}\right)=\frac{1}{2}\bm{n}\left(\bm{r}\right)\times\partial_{\mu}\bm{n}\left(\bm{r}\right)+\left(\bm{\tilde{A}}_{\mu}\left(\bm{r}\right)\cdot\bm{n}\left(\bm{r}\right)\right)
\bm{n}\left(\bm{r}\right),
\end{align*}
or equivalently,\begin{gather}
\bm{\hat{A}}_{\mu}\left(\bm{r}\right)=\frac{1}{2}\hat{\bm{z}}\times\left(\hat{\mathcal{R}}^{-1}\left(\bm{r}\right)\partial_{\mu}\bm{n}\left(\bm{r}\right)\right)
\nonumber\\
+
\left(\bm{\hat{A}}_{\mu}\left(\bm{r}\right)\cdot\hat{\bm{z}}\right)\hat{\bm{z}}.
\label{eq:covariant}
\end{gather}
The second term does not admit a covariant expression in terms of $\bm{n}\left(\bm{r}\right)$, reflecting the U(1) gauge ambiguity of $\hat{A}_{\mu}\left(\bm{r}\right)$. This is the gauge field associated with the Berry connection. The first term, however, is fully determined by the geometry of the crystal. This is the term associated with the affine connection.

\subsubsection{Affine connection term}

The field associated with the affine connection of the corrugated crystal is defined as\begin{align}
\hat{A}_{\mu}^{\textrm{aff}}\left(\bm{r}\right)=\frac{1}{2}\hat{\bm{z}}\times\left(\hat{\mathcal{R}}^{-1}\left(\bm{r}\right)\partial_{\mu}\bm{n}\left(\bm{r}\right)\right)\cdot\hat{\bm{s}}.
\end{align}
From the Weingarten equations~\eqref{eq:W1}-\eqref{eq:W2}, we have \begin{align}
\partial_{\mu}\bm{n}\left(\bm{r}\right)=-f_{\mu\nu}\left(\bm{r}\right)\bm{e}_{\nu}\left(\bm{r}\right).
\end{align}
This together with Eq.~\eqref{eq:Re} leads to\begin{align}
\hat{A}_{\mu}^{\textrm{aff}}\left(\bm{r}\right)=-\frac{1}{2}f_{\mu\nu}\left(\bm{r}\right)\epsilon_{\nu\lambda}\hat{s}_{\lambda}=\frac{i}{4}f_{\mu\nu}\left(\bm{r}\right)[\hat{s}_z,\hat{s}_{\nu}],
\end{align}
which is the result in Eq.~(5) of the main text.

The condition of smooth corrugations expressed in Eq.~(6) of the main text is deduced from an expansion to the lowest order in the displacements, see Eq.~\eqref{eq:expansion}, and from comparing this spin-lattice coupling with the spin-orbit splitting term in the Hamiltonian.

\subsubsection{Berry connection field}

We have to fix the gauge in order to evaluate the Berry connection term. We consider then the following unitary transformation,\begin{align}
 \hat{\mathcal{U}}\left(\bm{r}\right)=\left(\begin{array}{cc}
 \cos\frac{\theta\left(\bm{r}\right)}{2}e^{-i\phi\left(\bm{r}\right)/2} & -\sin\frac{\theta\left(\bm{r}\right)}{2}e^{-i\phi\left(\bm{r}\right)/2}\\
  \sin\frac{\theta\left(\bm{r}\right)}{2}e^{i\phi\left(\bm{r}\right)/2} & \cos\frac{\theta\left(\bm{r}\right)}{2}e^{i\phi\left(\bm{r}\right)/2}
 \end{array}\right),
 \end{align}
where we are parametrizing the normal to the surface according to a spherical representation:
\begin{align*}
\bm{n}\left(\bm{r}\right)=\left(\sin\theta\left(\bm{r}\right)\cos\phi\left(\bm{r}\right),\sin\theta\left(\bm{r}\right)\sin\phi\left(\bm{r}\right),\cos\theta\left(\bm{r}\right)\right).
\end{align*}
The associated SO(3) rotation reads \begin{widetext} \begin{align}
\hat{\mathcal{R}}\left(\bm{r}\right)=\left(\begin{array}{ccc}
 \cos\theta\left(\bm{r}\right)\cos\phi\left(\bm{r}\right) & -\sin\phi\left(\bm{r}\right)& \sin\theta\left(\bm{r}\right)\cos\phi\left(\bm{r}\right)\\
\cos\theta\left(\bm{r}\right)\sin\phi\left(\bm{r}\right) & \cos\phi\left(\bm{r}\right) & \sin\theta\left(\bm{r}\right) \sin\phi\left(\bm{r}\right)\\
-\sin\theta\left(\bm{r}\right) & 0 & \cos\theta\left(\bm{r}\right)
\end{array}\right).
\end{align}

From this and the definition in Eq.~\eqref{eq:A_def} we obtain\begin{gather*}
\hat{A}_{\mu}=\frac{1}{2}\left(-\sin\theta\left(\bm{r}\right)\partial_{\mu}\phi\left(\bm{r}\right) \hat{s}_x+\partial_{\mu}\theta\left(\bm{r}\right) \hat{s}_y+\cos\theta\left(\bm{r}\right)\partial_{\mu}\phi\left(\bm{r}\right) \hat{s}_z\right).
\end{gather*}
Note that the $\hat{s}_x$, $\hat{s}_y$ components can be obtained from the covariant expressions in Eq.~\eqref{eq:covariant}. The Berry connection field reduces to\begin{align}
A_{\mu}^{\textrm{B}}\left(\bm{r}\right)=\frac{1}{2}\cos\theta\left(\bm{r}\right)\partial_{\mu}\phi\left(\bm{r}\right).
\end{align}

The associated magnetic field is just\begin{align}
\mathcal{B}\left(\bm{r}\right)=\pm\frac{\hbar}{2e}\epsilon_{\mu\nu}\partial_{\mu}\left(\cos\theta\left(\bm{r}\right)\partial_{\nu}\phi\left(\bm{r}\right)\right)=\mp\frac{\hbar}{2e}\epsilon_{\mu\nu}\sin\theta\left(\bm{r}\right)\partial_{\mu}\theta\left(\bm{r}\right)\partial_{\nu}\phi\left(\bm{r}\right).
\end{align}
This field is gauge-invariant, and therefore admits a covariant expression in terms of $\bm{n}\left(\bm{r}\right)$. By noticing that \begin{align}
\sin\theta\left(\bm{r}\right)\left(\partial_{\mu}\theta\left(\bm{r}\right)\partial_{\nu}\phi\left(\bm{r}\right)-\partial_{\nu}\theta\left(\bm{r}\right)\partial_{\mu}\phi\left(\bm{r}\right)\right)=\bm{n}\left(\bm{r}\right)\cdot\left(\partial_{\mu}\bm{n}\left(\bm{r}\right)\times\partial_{\nu}\bm{n}\left(\bm{r}\right)\right),
\end{align}
we arrive at\begin{gather}
\mathcal{B}\left(\bm{r}\right)=-\frac{\hbar}{4e}\epsilon_{\mu\nu}\,\bm{n}\left(\bm{r}\right)\cdot\left(\partial_{\mu}\bm{n}\left(\bm{r}\right)\times\partial_{\nu}\bm{n}\left(\bm{r}\right)\right).
\label{eq:E}
\end{gather}\end{widetext}
The last expression can be rewritten as
\begin{align}
\bm{n}\left(\bm{r}\right)\cdot\left(\partial_{\mu}\bm{n}\left(\bm{r}\right)\times\partial_{\nu}\bm{n}\left(\bm{r}\right)\right)=
\epsilon_{\lambda\rho}S_{\lambda\mu}\left(\bm{r}\right)S_{\rho\nu}\left(\bm{r}\right),
\end{align}
just by using Weingarten equations~\eqref{eq:W1}. At the same time, we have
\begin{align}
\epsilon_{\mu\nu}\epsilon_{\lambda\rho}S_{\lambda\mu}\left(\bm{r}\right)S_{\rho\nu}\left(\bm{r}\right)=2\,\text{det}\,\hat{S}\left(\bm{r}\right)=2\kappa\left(\bm{r}\right),
\end{align}
from which Eq.~(7) of the main text is deduced immediately.

\subsection{Magnetic flux and topological defects}

\subsubsection{Global Gauss-Bonnet theorem}

For any compact orientable surface $S$ with smooth boundary $\partial S$ the following identity holds
\begin{equation}
\int_{S}\kappa d A+\int_{\partial S}\kappa_{g}ds=2\pi\chi\left(S\right),
\end{equation}
where $s$ and $\kappa_{g}$ are the arc-length parametrization and the geodesic curvature of the boundary, respectively. The geodesic curvature measures how far a curve is from being a geodesic of the surface.

The Euler characteristic is an integer-valued topological index characterizing the global shape of the surface. In the case of connected surfaces, the Euler characteristic becomes $\chi(S)=2(1-g)$, where $g$ is its genus (number of 'holes'). 

\subsubsection{Topological charge of defects from Volterra's construction}

\begin{figure}[t!]
\begin{center}
%\hspace{-0.4cm}
\includegraphics[width=6cm]{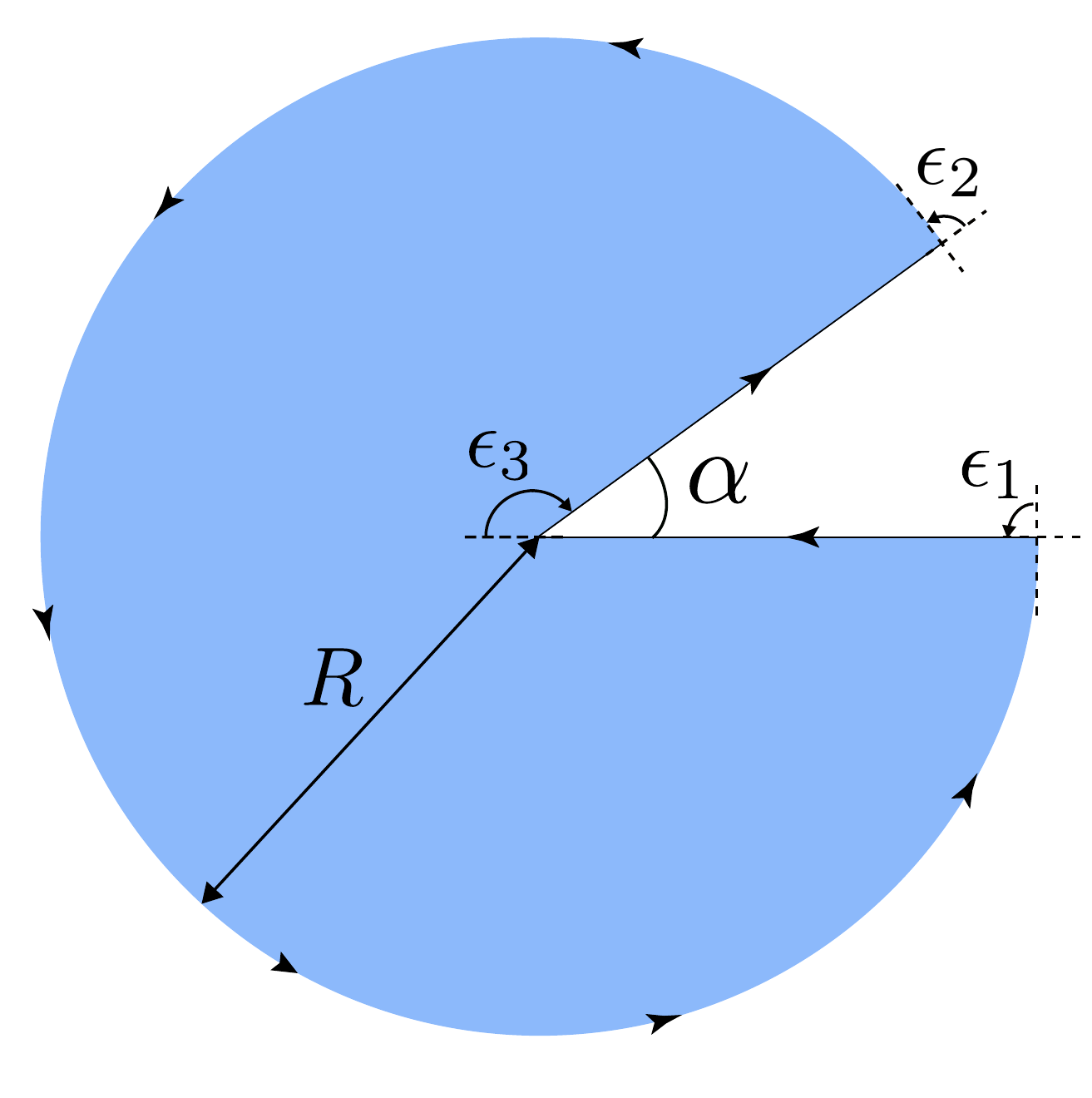}
%\vspace{-4.1cm} 
\caption{Lattice without the angular sector (before identification) in the continuum limit. The sum of the exterior angles equals $\alpha$.}
\label{Fig2}
\end{center}
\end{figure}

Disclinations can be constructed by removing or adding an angular sector of the material and then by identifying the boundaries. Consider a disk $D$ centered at the origin with radius $R$. We can apply the local Gauss-Bonnet theorem, which reads
%Fig. \ref{Fig2} shows the Volterra's representation $D$ of the disclination in the $xy$ plane, which is characterized by the removal of an angular sector of angle $\alpha$. The excess angle is $\alpha=\pm\frac{2\pi}{3}$ in the case of tetragonal/octogonal disclinations. The total curvature of the disclination can be calculated via the local Gauss-Bonnet theorem
%The Gaussian curvature has two contributions in this planar representation of the surface, one bulk denoted by $K$ and another resulting from the geodesic curvature $\kappa_{g}$ associated with the boundary. The total curvature can be calculated via the local Gauss-Bonnet theorem 
\begin{equation}
\int_{D}\kappa\, \textrm{d} A+\int_{\partial D}\kappa_{g}\textrm{d}s+\sum_{i}\epsilon_{i}=2\pi.
\end{equation}
Here $\{\epsilon_{i}\}_{i}$ are the exterior angles of the (piecewise) smooth boundary $\partial D$.

Before the removal, we have $\kappa=0$, $\sum\epsilon_{i}=0$, and $\kappa_g=R^{-1}$. When a sector of angle $\alpha$ is removed, the integral of the geodesic curvature is $2\pi-\alpha$. This loss is compensated with the contribution from the exterior angles, see Fig. \ref{Fig2}, whose sum equals the defect angle $\alpha$. When the boundaries are identified, the external angles disappear and the loss of geodesic curvature is compensated with the appearance of Gaussian curvature at the cusp, $\kappa\left(\bm{r}\right)=\alpha\,\delta^{(2)}\left(\bm{r}\right)$. Similarly, when an angular sector is added the excess of geodesic curvature at the boundary must be compensated with negative Gaussian curvature at the cusp. The same result can be obtained by a direct calculation of the curvature from the metric of the conical-like surface. The defect angle $\alpha$ can then be identified with the topological charge of the defect. 

\subsubsection{Euler's formula and Dirac quantization condition}

For any (convex) polyhedron surface $S$ the Euler characteristic becomes $\chi(S)=F-E+V=2$, where $F$, $E$ and $V$ denote the number of faces, edges and vertices, respectively. In this Letter, the surface considered consists of an hexagonal lattice where topological defects (disclinations) are embedded. Let $n_{k}$, $k\geq3$, denote the number of $k$-polygons being part of the polyhedron surface ($k=3$ corresponds to triangles, $k=4$ to tetragons, etc). 

In the case of a closed surface, the coordination number of the edges and vertices is two and three, respectively. This means that each edge (vertex)  results from the intersection of only two (three) polygons. Therefore, we have \begin{gather*}
\displaystyle F=\sum_{k\geq3}n_{k}, \\
\displaystyle E=\frac{1}{2}\sum_{k\geq3} k\, n_{k},\\
\displaystyle V=\frac{1}{3}\sum_{k\geq3} k\, n_{k},
\end{gather*}
and the Euler's formula becomes
\begin{equation}
12=3n_{3}+2n_{4}+n_{5}-n_{7}-2n_{8}-\ldots
\end{equation} 

In the particular case of TMDC crystals, we only consider tetragonal and octogonal defects. The above formula becomes\begin{equation}
6=n_{4}-n_{8}.
\end{equation} 

%\subsection{Spin Hall effect due to topological defects: Drude model}

\subsection{Aharonov-Bohm effect: diagrammatic calculation}

\begin{figure}[t!]
\begin{center}
\hspace{-0.4cm}
\includegraphics[width=6.5cm]{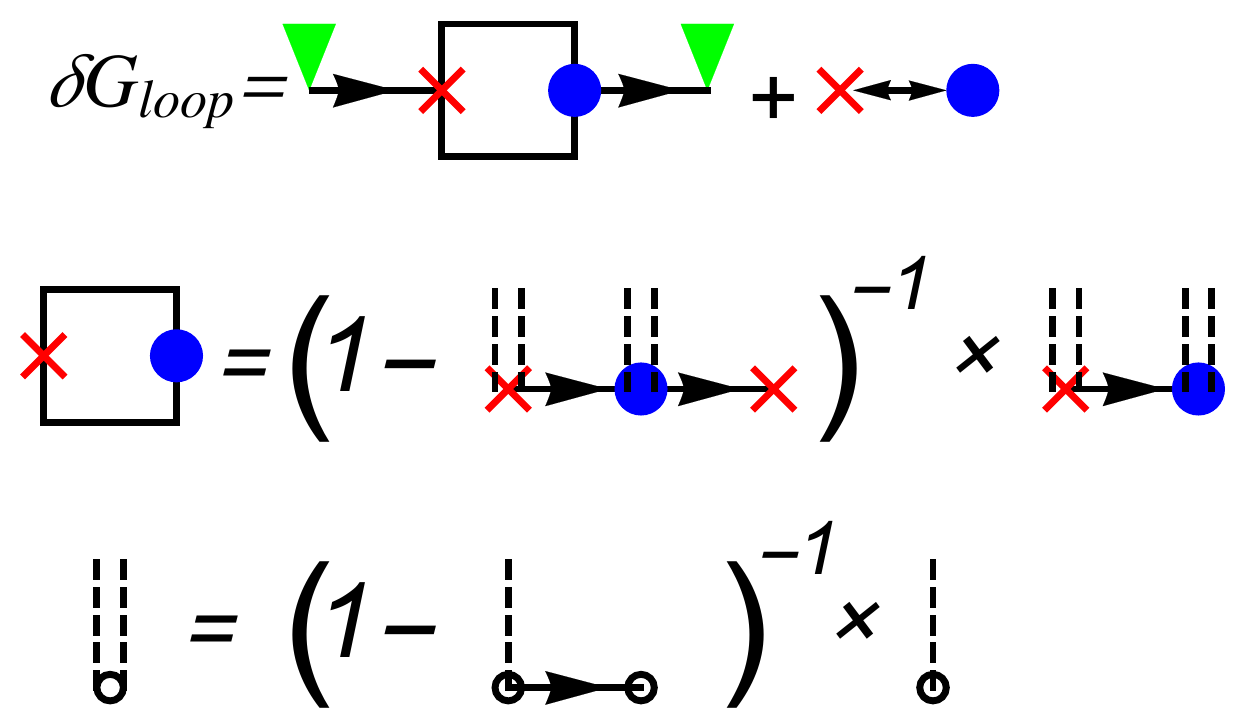}
%\vspace{-3.1cm} 
\caption{Feynman diagrams for the calculation of $\delta G_{\textrm{loop}}\left(\bm{r},\bm{r},\omega\right)$. The black arrows represent the retarded Green function defined in Eq.~\eqref{eq:G0}. The green arrow represents the position of the STM tip, $\bm{r}$, the red cross labels the position of the first impurity, $\bm{r}_1$, and the blue dot is the position of the second impurity, $\bm{r}_2$. The box represents the effective non-local potential $V$ arising from scattering back and forth the two impurities. The single dashed line represents the bare scattering potential $U_0$, and the double dashed line is the dressed scattering potential $U$ in the T-matrix approximation.}
\label{diagram}
\end{center}
\end{figure}

We consider the following Hamiltonian for the interferometer discussed in the main text:
\begin{align}
\mathcal{H}=\mathcal{H}_0+\sum_{i=1,2}U_0\,\delta^{(2)}\left(\bm{r}-\bm{r}_i\right).
\end{align}
Here $\bm{r}_{1,2}$ label the positions of the scattering centers and $U_0$ is the strength of the scattering potential. We neglect spin relaxation and inter-valley scattering, so the contributions from different carrier species sum up and we can therefore take just 
\begin{align}
\mathcal{H}_0=\frac{\hbar^2\bm{k}^2}{2m^{*}}.
\end{align}

\begin{center}
\begin{table*}
\begin{tabular}{|c|c|c|}
\hline
Relaxation gaps & Relaxation rates & Contribution to $\Delta\sigma_{xx}\left(B\right)$\\
\hline
\hline
$\Gamma_0^0=0$ & Gapless (time-reversal symmetry) & Negative \\
%\hline
$\Gamma_x^0,\Gamma_y^0\rightarrow\infty$ & Gapful & Absent \\
%\hline
$\Gamma_z^0$ & $2\tau_{sf}^{-1}+2\tau_{is}^{-1}$ & Positive \\
%\hline
$\Gamma_0^x,\Gamma_0^y$
& Gapful & Absent\\
%\hline
$\Gamma_x^x,\Gamma_x^y,\Gamma_y^x,\Gamma_y^y$ &
$\tau_{\star}^{-1}+
\tau_{sf}^{-1}+\tau_{is}^{-1}$ & Positive\\
%\hline
$\Gamma_z^x,\Gamma_z^y\rightarrow\infty$ & Gapful & Absent\\
%\hline
$\Gamma_0^z$&
$2\tau_{is}^{-1}$ & Positive\\
%\hline
$\Gamma_x^z,\Gamma_y^z\rightarrow\infty$ & Gapful & Absent\\
%\hline
$\Gamma_z^z$&
$2\tau_{sf}^{-1}$ & Negative\\
\hline
\end{tabular}
\caption{Relation between Cooperon relaxation gaps and scattering rates.}
\label{gaps}
\end{table*}
\end{center}

As deduced from the diagrams in Fig.~\ref{diagram}~(b), the correction to the local Green function due to semiclassical loop trajectories can be expressed as
\begin{align*}
\delta G_{\textrm{loop}}\left(\bm{r},\bm{r},\omega\right)=G_{0}\left(\bm{r}-\bm{r}_2,\omega\right)V\left(\bm{r}_2,\bm{r}_1,\omega\right)G_{0}\left(\bm{r}_1-\bm{r},\omega\right)
\\
+\,\,\bm{r}_1\leftrightarrow\bm{r}_2,
\end{align*}
where the $G_0$ is the (retarded) Green function in the absence of scattering centers, 
\begin{align}
\label{eq:G0}
G_{0}\left(\bm{r}-\bm{r}',\omega\right)\equiv \left\langle\bm{r}\right|\left(\omega+i\,0^+-\hat{\mathcal{H}}_0\right)^{-1}\left|\bm{r}'\right\rangle,
\end{align}
and $V$ -- the box in the diagrams -- can be interpreted as an effective non-local potential arising from trajectories back and forth the two scattering centers:\begin{widetext}
\begin{gather}
V\left(\bm{r}_2,\bm{r}_1,\omega\right)=U\left(\omega\right)G_0\left(\bm{r}_2-\bm{r}_1,\omega\right)U\left(\omega\right)+U\left(\omega\right)G_0\left(\bm{r}_2-\bm{r}_1,\omega\right)U\left(\omega\right)G_0\left(\bm{r}_1-\bm{r}_2,\omega\right)U\left(\omega\right)G_0\left(\bm{r}_2-\bm{r}_1,\omega\right)U\left(\omega\right)+\ldots\nonumber\\
=U\left(\omega\right)G_0\left(\bm{r}_2-\bm{r}_1,\omega\right)U\left(\omega\right)+U\left(\omega\right)G_0\left(\bm{r}_2-\bm{r}_1,\omega\right)U\left(\omega\right)G_0\left(\bm{r}_1-\bm{r}_2,\omega\right)V\left(\bm{r}_2-\bm{r}_1,\omega\right).
\end{gather}
Here $U\left(\omega\right)$ is the dressed scattering potential in the T-matrix approximation, accounting for multiple scattering off the same impurity,\begin{align}
U\left(\omega\right)=U_0+U_0G_0\left(0,\omega\right)U_0+U_0G_0\left(0,\omega\right)U_0G_0\left(0,\omega\right)U_0+\ldots=U_0+U_0G_0\left(0,\omega\right)U.
\end{align}\end{widetext}
Notice that the real part of $G_0\left(\bm{r},\omega\right)$ is singular at the origin since $G_0\left(\bm{r},\omega\right)\sim H_0^{(1)}\left(|\bm{r}|\sqrt{\frac{2m^{*}\omega}{\hbar}}\right)$, where $H_0^{(1)}\left(x\right)$ is a Hankel function. It must be regularized by introducing a bandwidth $\Lambda$. In the calculation of Fig.~3 of the main text we have taken $\Lambda=10\times\frac{\hbar^2k_F^2}{2m^{*}}$. The rest of parameters are specified in the caption.

Finally, in the derivation of the final expression for $\delta\rho$ in the main text we employed the semiclassical approximation for the Green function,
\begin{align*}
G\left(\bm{r}_1,\bm{r}_2,\omega\right)=\text{Exp}\left[i\int_{\bm{r}_1}^{\bm{r}_2}d\bm{l}\cdot\bm{A}^{\textrm{B}}(\bm{r})\right]G_0\left(\bm{r}_1,\bm{r}_2,\omega\right).
\end{align*}

\subsection{Weak (anti-)localization and magneto-conductance}

In TMDCs, the quantum interference correction to the conductivity is given by 16 
"Cooperon" modes $\tilde{C}_s^{l}$ labeled by spin $s=0,x,y,z$ and valley $l=0,x,y,z$ indices,
\begin{gather}
\delta \sigma_{xx}=-\frac{2e^2D}{\pi\hbar}\int
\frac{d^2\mathbf{Q}}{\left(2\pi\right)^2}\sum_{s,l}c_sc^l\tilde{C}_s^l\left(\mathbf{Q},\omega=0\right),
\mbox{ with}
\nonumber\\
c_{0,x,y,z}=-1,+1,+1,+1, \nonumber\\c^{0,x,y,z}=+1,+1,+1,-1.
\label{eq:WL2}
\end{gather}
Here $D=v_F^2\tau/2$ is the diffusion coefficient, with $v_F$ and $\tau$ being the Fermi velocity and total scattering time, respectively. The integral in $\mathbf{Q}$ is cut-off by the inelastic coherence length $l_{\varphi}^{-1}$ in the infrared.

The Cooperons are two-particle correlation functions obeying a Bethe-Salpeter equation. Due to the spin-orbit splitting of the bands, only two groups of 4 Cooperons corresponding to singlet ($s,l=0$) and triplet ($s,l=x,y,z$) combinations built separately of two Kramers doublets ($K_+,\uparrow$; $K_-,\downarrow$) and ($K_+,\downarrow$; $K_-,\uparrow$) are gapless in the limit of purely potential disorder. In the diffusive approximation, $\ell\left|\mathbf{Q}\right|,\tau\omega\ll 1$, these 8 modes become solutions of the diffusion-relaxation kernels\begin{align}
\left[-D\nabla^2-i\omega+\Gamma_s^{l}\right]\tilde{C}_s^{l}\left(\mathbf{r}-\mathbf{r}',\omega\right)=\delta\left(\mathbf{r}-\mathbf{r}'\right).
\end{align}

The relation between the relaxation gaps $\Gamma_{s}^l$ and the different scattering channels are studied in detail in Ref.~29 of the main text. Here for simplicity we consider $\tau^{-1}=\tau_{0}^{-1}+\tau_{sf}^{-1}+\tau_{is}^{-1}+\tau_{\star}^{-1}$, where $\tau_{\alpha}$ are the scattering times related to the following elastic channels: $\tau_0$ is the scattering time associated with potential disorder, which is assumed to dominate over the rest; $\tau_{sf}$ and $\tau_{is}$ are the intra-valley and inter-valley spin relaxation times, respectively; finally, $\tau_{\star}$ is the scattering time associated with ripples or topological defects, determining the dephasing lengths limited by the Gaussian curvature of the sample. These rates enter in the Cooperon relaxation gaps as indicated in Tab.~\ref{gaps}. The magneto-conductivity reads in general
\begin{gather}
\Delta \sigma_{xx}\left(B\right)=\frac{e^2}{\pi h}\left[4F\left(\frac{B}{B_{\varphi}+B_{sf}+B_{is}+B_{\star}}\right)\right.\nonumber\\\left.
+F\left(\frac{B}{B_{\varphi}+2B_{sf}+2B_{is}}\right)
+F\left(\frac{B}{B_{\varphi}+B_{is}}\right)
\right.\nonumber\\\left.
-F\left(\frac{B}{B_{\varphi}+B_{sf}}\right)-F\left(\frac{B}{B_{\varphi}}\right)
\right],
\label{eq:delta_sigma}
\end{gather}
where $F\left(x\right)$, and $B_{\alpha}$ are defined as in the main text. 

The total scattering time reads $\tau_{s}^{-1}=\tau_{sf}^{-1}+\tau_{is}^{-1}$. It is natural to assume that both times are of the same order, $\tau_{sf}\sim\tau_{is}$, since spin diffusion is likely to be determined by vacancies in the layer of chalcogen atoms, which break the mirror symmetry --and therefore inducing spin relaxation-- and induce short-ranged potentials that are able to mix both valleys. Under that assumption, the magneto-conductance formula is simplified to the one shown in Eq.~(10) of the main text. As discussed there, the resulting magneto-conductance is strongly suppressed when $l_{\star}\ll l_{\varphi}$, becoming negative at low fields when\begin{align}
l_{s}<\frac{l_{\varphi}^2}{\sqrt{2} \, l_{\star}}.
\end{align}

We estimate now the magnitude of $l_{\star}$. The phase acquired by a quasiparticle describing a loop of length $L\sim R$ close to the center of a ripple is of the order of $\varphi\sim\left(h_0/R\right)^2$, where $h_0$ is the typical height and $R$ is the lateral size. In the case of an edge dislocation --one tetragonal and one octogonal disclinations separated by a distance $d$--, the phase is 0 when the loop encloses completely the dislocation, but equals half the topological charge of one of the disclinations, $\varphi=\alpha/2=\pm\pi/3$, when the trajectory is near the core of one of them. By taking into account that the width of the semi-classical trajectories is of the order of $k_F^{-1}$, with $k_F$ being the Fermi wave vector, the number of defects contributing to the dephasing of the wave function can be estimated as
\begin{align}
N\sim\begin{cases}
\frac{k_F^{-1}L}{R^2}&\text{in the case of ripples},\\
\frac{k_F^{-1}L}{d^2}&\text{in the case of dislocations}.
\end{cases}
\end{align}
Since the process is completely random, from the central limit theorem we deduce that the dephasing due to $N$ defects is just $\varphi_N\sim\sqrt{N}\varphi$. The dephasing length can be defined as the length of the loops $L=l_{\star}$ for which $\varphi_{N}\sim 1$, leading to
\begin{align}
l_{\star}\sim\begin{cases}
\frac{k_FR^6}{h_0^4}&\text{in the case of ripples},\\
k_F d^2&\text{in the case of dislocations}.
\end{cases}
\end{align}

Close to the band edge, when the condition for the adiabatic regime is fulfilled, $l_{\star}$ is very short and a strong suppression of the localization effects is expected.


\begin{thebibliography}{99}
\bibitem{spin-Hall} Analogously to the anomalous Hall effect in ferromagnets, a dilute concentration $\rho$ of disclinations with charge $\alpha$ engenders an additional term in the spin-Hall conductivity, $\sigma_{xy}^{\textrm{sH}}\approx\frac{\hbar\tau\sigma_{xx}}{2 m^{*}}\alpha\rho$, where $\sigma_{xx}=\frac{e^2 n\tau}{m^{*}}$ is the Drude conductivity, $n$ is the carrier concentration, and $\tau$ is the scattering time.
\end{thebibliography}
\end{document}